\documentclass[preprint,5p]{elsarticle}
\usepackage{epsfig}
\everymath{\displaystyle}
\usepackage{dcolumn}

\def\as{\alpha_{\rm s}}
\newcommand{\beg}{\begin{equation}}
\newcommand{\en}{\end{equation}}

\def\albars{\relax\ifmmode{\bar{\alpha}_s}
    \else{$\bar{\alpha}_s${ }}\fi}
 \def\msbar{\relax\ifmmode\overline{\rm MS}
     \else{$\overline{\rm MS}${ }}\fi}

\begin{document}

\title{Four-loop QCD analysis of the Bjorken sum rule vs data}

\author{
V.L.~Khandramai$^a$, R.S.~Pasechnik$^b$, D.V.~Shirkov$^{c}$,
O.P.~Solovtsova$^{a,c}$, O.V.~Teryaev$^{c}$
 \\ [2mm]
{}$^a${\small Gomel State Technical University, 246746 Gomel,
Belarus}
\\
{}$^b${\small High Energy Physics, Department of Physics and
Astronomy, Uppsala University,} \\
{\small SE-75121 Uppsala, Sweden}
\\
{}$^c${\small Bogoliubov Laboratory of Theoretical Physics, Joint
Institute for Nuclear Research,} \\
{\small 141980 Dubna, Moscow Region, Russia} }
\date{\today}
\begin{abstract}

 We study the polarized Bjorken sum rule at low momentum transfers in
 the range $0.22<Q<1.73 ~{\rm GeV}$ with the four-loop N$^3$LO
 expression for the coefficient function $C_{\rm Bj}(\as)$ in the
 framework of the common QCD perturbation theory (PT) and the
 singularity-free analytic perturbation theory (APT). The analysis of
 the PT series for $C_{\rm Bj}(\as)$ gives a hint to its asymptotic
 nature manifesting itself in the region $Q<1$ GeV. It relates to the
 observation that the accuracy of both the three- and four-loop PT
 predictions happens to be at the same 10\% level. On the other hand,
 the usage of the two-loop  APT allows one to describe the precise
 low energy JLab data down to $Q\sim 300$ MeV and gives a possibility
 for reliable extraction of the higher twist (HT) corrections.  At
 the same time, above $Q\sim 700$ MeV the APT two-loop order with HT
 is equivalent to the four-loop PT with HT compatible to zero and is
 adequate to current accuracy of the data.


\end{abstract}


\maketitle

\newpage
\section{Introduction}

The higher order perturbative QCD (pQCD) and higher twist
corrections become very important, in particular, in observables of
the Deep-Inelastic Scattering (DIS) at low momentum transfers $Q\leq
1$ GeV. The most precise low-energy data from the Jefferson Lab
\cite{JLab08data,JLab-old-data} on one of the main sources of
information about the nucleon structure, the Bjorken sum rule (BSR)
\cite{Bj}, are the real challenge to the accuracy of the pQCD
expansions. In our previous papers \cite{Bjour,Pasechnik:2010fg}, we
studied this issue at the three-loop level. In the current paper, we
continue this line of investigations and explore the effect of the
recent calculation \cite{Baikov:2010} of the four-loop (in $\as$)
contribution to the BSR.

The BSR claims that the difference of
the proton and neutron structure functions integrated over all
possible values
\begin{equation}\label{Bj}
\Gamma^{p-n}_1(Q^2) \,=\, \int_0^{1}\,\ \left[g_1^p(x,Q^2) -
g_1^n(x,Q^2) \right] dx \, , \end{equation} of the Bjorken variable
$x$ in the limit of large four-momentum squared of the exchanged
virtual photon, $Q^2 \to \infty $, is equal to $g_A/6$, with
$g_A=1.267\pm0.004$~\cite{PDG10}, the nucleon axial charge defined
from the neutron $\beta$-decay data.

 The r.h.s. of Eq.~(\ref{Bj}) is given by a sum of two series in powers
 of $1/Q^2$ (OPE higher twists corrections) and in powers of the QCD
 running coupling $\as(Q^2)$ (pQCD radiative corrections). Until very
 recently, the pQCD contribution to BSR was known ~\cite{LV91} up to
 the third order $\sim\as^3$. So far, the corresponding expression
 has been used in many studies, in particular, for extraction of the
 $\as$ values at low momentum scales ~\cite{EK95}.

 One of the actual theoretical subjects is the interplay between the
 higher twists (HT) and higher order pQCD corrections at low $Q$,
 which has recently been studied in Refs.~\cite{Bjour} at the
 three-loop level. There, it was shown that the satisfactory
 description of the data down to $Q_{min}\sim \Lambda_{\rm
 {QCD}}\simeq 350$ MeV can be achieved within the Analytic
 Perturbation Theory (APT), the ghost-free modification of pQCD. In
 the current work  we repeat this analysis at the four-loop N$^3$LO
 level.

The APT approach is based on the causality principle implemented as
the analyticity imperative in the complex $Q^2$-plane for the QCD
coupling $\as(Q^2)\,$ in the form of the K\"allen-Lehmann spectral
representation~\cite{apt96-7} and on the demand of compatibility with
linear integral transformations \cite{nonpol99} (for an overview on
the APT concept and results, see~Ref.~\cite{Sh-revs}). It is
well-known that in the APT framework, the theoretical ambiguity
associated with pQCD higher-loop corrections is diminished (see
Ref.~\cite{apt-DV-2001}), and results are practically renormalization
scheme independent \cite{MSS_PL_97}.

The four-loop expression for the pQCD contribution to the Bjorken sum
rule became recently available in Ref.~\cite{Baikov:2010}. It gives
us a reasonable motivation for a new extended QCD analysis of the
combined JLab data on $\Gamma_1^{p-n}(Q^2)$ at low $0.05<Q^2<3.0$
GeV$^2$ accounting for up to $\as^4$-order in both the (standard) PT
and APT approaches.

The paper is organized as follows. In Section 2, we study the higher
loop stability of both the PT and APT series and the renormalisation
scale dependence of the higher-order PT expansion for the Bjorken
sum rule. Section 3 contains the QCD results on extraction of the
higher twist terms from the experimental data at the four-loop
level. Summarizing comments are given in the last section.

\section{The perturbative QCD contribution}

Commonly, one represents the Bjorken integral (\ref{Bj}) as a sum of
the perturbative and the higher twist contributions
\begin{eqnarray} \label{PT-Bj-HT}
\Gamma^{p-n}_1(Q^2)=\frac{g_A}{6}\biggl[1-\Delta_{\rm Bj}(Q^2)
\biggr]+\sum_{i=2}^{\infty}\frac{\mu_{2i}}{Q^{2i-2}} \,.
\end{eqnarray}
The perturbative QCD correction\footnote{This correction is
defined by the coefficient function, $\Delta_{\rm Bj}=1-C_{\rm
Bj}(\as)$.} $\Delta_{\rm Bj}(Q^2)$ has a form of the power series in
the QCD running coupling $\as(Q^2)$. At the up-to-date four-loop
(N$^3$LO) level in the massless case it looks like
\begin{equation}\label{Delta_PT-Bj}
 \hspace{17mm} \Delta_{\rm Bj}^{\rm PT}(Q^2) =
 \sum_{k \leq 4}\,\, c_k \,\as^k(Q^2)\,.
\end{equation}
Here, the numerical expansion coefficients $c_i$ in the modified
minimal subtraction ($\overline{\rm{MS}}$) scheme, for three active
flavors, $n_f=3$, read $c_1=1/\pi=0.31831$,
$c_2=0.36307$~\cite{GL:86}, $c_3=0.65197$~\cite{LV91} and
$c_4=1.8042$ \cite{Baikov:2010}. Besides, the four-loop running
coupling $\as(Q^2)$ is defined as a solution of the Renormalization
Group (RG) equation
\begin{equation}\label{rg4loop}
  \frac{d\as}{dL}= \beta(\as)\,;\quad
  \beta(\as) =
  \sum_{0\leq k \leq 3}\,\,\beta_k\,\as^{k+2}\,,
\end{equation}
where $L=\ln(\mu^2/\Lambda^2)$ and $\beta_k$ are the coefficients of
the $\beta$-function. For our purposes, it is convenient to represent
the $\beta$-function in the form
\begin{equation}
 \beta(\as)\,=-\beta_0\,\as^2\,
 (1+b_1\as+b_2\as^2+b_3\as^3+\ldots),
\end{equation}
with $b_i=\beta_i/\beta_0$, the ratios of the $\beta$-function
coefficients. For three flavors the coefficients are
$\beta_0=9/4\pi=0.7162$, $b_1= 0.5659$,
$b_2^{\overline{\rm{MS}}}= 0.4530$ \cite{beta-3} and
$b_3^{\overline{\rm{MS}}}= 0.6770$ \cite{beta-4}.
In the current analysis we use the exact solutions of the RG equation
(\ref{rg4loop}) in the $\overline{\rm{MS}}$-scheme at the scale
$\mu=Q$.

\subsection{Analytic Perturbation Theory}

The moments of the structure functions are analytic functions in the
complex $Q^2$-plane with a cut along the negative part of the real
axis (see, e.g., Ref.~\cite{JLD-00}).
The perturbative representation (\ref{Delta_PT-Bj}) violates these
analytic properties due to the unphysical singularities of $\as(Q^2)$
for $Q^2>0$. To resolve the issue, we apply the APT
method~\cite{apt96-7,Sh-revs}, which allows one to combine the RG
invariance with proper analytical properties of the RG-invariant
coupling and observables. In particular, the four-loop APT expansion
for the perturbative part $\Delta_{\rm Bj}(Q^2)$ is given by
\begin{equation}
\label{Delta_APT-Bj}
 \Delta_{\rm Bj}^{\rm APT}(Q^2)= \sum_{k
 \leq 4}\,\, c_k\,{\cal{A}}_{k}(Q^2)\,.
\end{equation}
Here the coefficients $c_k$ are the same as in Eq.~(\ref{Delta_PT-Bj}),
and the functions ${\cal{A}}_{k}(Q^2)$ are defined through the
spectral functions $\varrho_k(\sigma)\equiv{\rm Im}\left[\as^{k}
(-\sigma-i\epsilon)\right]$ by the spectral integral
\begin{equation}\label{delta_n}
{\cal{A}}_{k}(Q^2)=\frac{1}{\pi}\int\limits_0^\infty\, d\sigma\,
\frac{\varrho_k (\sigma)}{\sigma+Q^2}\,.
\end{equation}
Note, the first function, ${\cal{A}}_{1}(Q^2)$, is the analytic
coupling, ${\alpha}_{\rm APT}(Q^2) \,=\, {\cal{A}}_{1}(Q^2)\,.$ At
large momentum transfers, all the functions ${\cal{A}}_{k}(Q^2)$
become proportional to the $k$-th power of the usual perturbative
coupling $[\as(Q^2)]^k$ and the expansion (\ref{Delta_APT-Bj})
reduces to the power series (\ref{Delta_PT-Bj}). However, at small
enough $Q\leq 1 - 2 ~{\rm GeV}$ the properties of the non-power
expansion (\ref{Delta_APT-Bj}) become considerably different from
the PT power series (\ref{Delta_PT-Bj}) (see, e.g.,
Ref.~\cite{MSS_PL_97} for details).

\subsection{The $Q^2$-dependence}

Now we analyze the $Q^2$-dependence of the BSR in the framework of
both the PT and APT approaches in different orders (NLO, N$^2$LO and
N$^3$LO) of the perturbative expansions~(\ref{Delta_PT-Bj}) and
(\ref{Delta_APT-Bj}), respectively.
As a normalization point, we use the most accurate $\as$-value at
$Q= M_Z$, $\as{(M_Z)}=0.1184\pm0.0007$ \cite{PDG10,Bethke:2009}. In
order to take into account flavor thresholds, we apply the matching
conditions for the values of $\as(Q^2)$ which are rather nontrivial in
higher PT orders (see Refs.~\cite{CheKniSte97,SrSt,Kotikov}).
Following to analysis in Ref.~\cite{rundec}, our matched calculation
for the four-loop $\overline{\rm{MS}}$-coupling gives
$\Lambda^{(n_f=3)}= 336 \pm 10$ MeV. Note, we obtain practically the
same results, but with larger errors, if we choose the
pseudo-observable value $R(M_Z^2)=1.03904\pm0.00087$ as a
normalization point \cite{Ch-tau}, which leads to the four-loop
running coupling equal to $\as(M_Z)=0.1190\pm0.0026$.

\begin{figure}[h!]
 \centerline{\epsfig{file=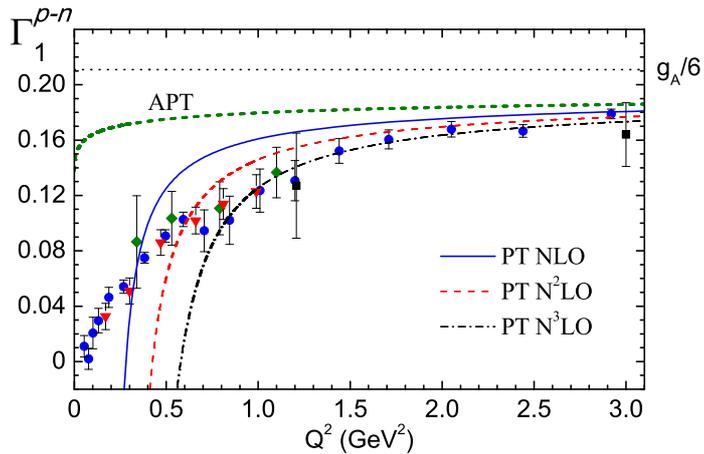,width=10.0cm}}
 \caption{Perturbative part of the BSR as a function of the momentum
 transfer squared $Q^2$ in different orders in both the APT
 and standard PT approaches against the combined set of
 the Jefferson Lab \cite{JLab08data,JLab-old-data} and SLAC \cite{SLAC} data.}
 \label{fig:PT_Gam}
\end{figure}

In Fig.~\ref{fig:PT_Gam}, we illustrate the behavior of the
perturbative part of the BSR in different orders in $\as$ in both PT
and APT approaches. The APT curves in different orders (NLO, N$^2$LO
and N$^3$LO) practically (at about $1~\%$ accuracy) coincide with
each other, so we represent the APT result by a single dash-dotted
line in Fig.~\ref{fig:PT_Gam}.
For completeness, we also show here the combined SLAC and JLab data
on $\Gamma_1^{p-n}(Q^2)$ used in our analysis. The SLAC data points
\cite{SLAC} are denoted by squares, the JLab CLAS Hall~A 2002 data
-- by downward pointing triangles, the JLab CLAS Hall~B 2003 data --
by diamonds \cite{JLab-old-data}, and the most recent JLab data
\cite{JLab08data} -- by circles. The horizontal dotted line
represents the limiting value $\Gamma_1^{p-n}(Q^2 \to
\infty)=g_A/6$.

One can see that at $Q^2 \geq 0.7$ GeV$^2$ the four-loop
approximation describes the data quite well. Moreover, the
corresponding curve passes close to the central values of several
data points, although the experimental accuracy (which is of the
same order as both the three- and four-loop contributions) does not
allow one to make a definite choice between four- and three-loop
approximations.

At the same time, at $Q^2 \leq 0.7$ GeV$^2$ the four-loop
approximation describes the data equally bad as the three- and
two-loop ones. This a signal of the necessity to account for HT
contributions, and it will be strongly dependent on the order of PT
used for its extraction \cite{Bjour}.

This  changes when APT is applied and the higher-loop stability is
achieved. This is a well-known feature of APT free from unphysical
singularities. At the same time, the deviation of APT curve from the
data shows for necessity of the HT contribution which in this case is
quite stable \cite{Bjour}.

This situation may be considered as an indication of the transition
of PT series to the asymptotic regime (while APT series remains
convergent) for $Q^2 \sim 0.7$ GeV$^2$. Let us explore this
possibility in more detail.

\subsection{Convergence of the PT and APT expansions}

Clearly, at low $Q^2$ a value of the strong coupling is quite large,
questioning the convergence of perturbative QCD series.
The PT power series truncated after four-loop
order (c.f. Eq.~(\ref{Delta_PT-Bj})) reads
\begin{eqnarray} \nonumber
& \Delta_{\rm Bj}^{\rm PT}(\as) = 0.3183 \, \as \, + \,0.3631 \,
\as^2
\\
&~~~~ ~~~~+ 0.6520 \,  \as^3 + \, 1.804 \, \as^4=  \sum_{i
 \leq 4}\,{\delta_i (\as)},
\label{Delta-ex}
\end{eqnarray}
where $\delta_i$ is the $i$-th term. The qualitative resemblance of
the coefficients pattern to the factorial growth did not escape our
attention although the more definite statements, if possible, would
require much more efforts. This observation allows one to estimate the
value of $\as \sim 1/3$ providing a similar magnitude of
three- and four- loop contributions to the BSR.
\begin{figure}[h!]
 \centerline{\epsfig{file=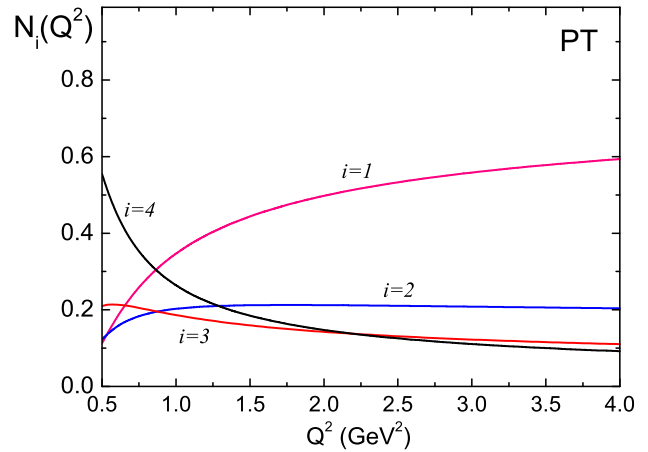,width=10cm}}
 \caption{The $Q^2$-dependence of the relative contributions at the four-loop
level in the PT approach. Four-loop PT order overshoots the
three-loop one at $Q^2\leq 2$ GeV$^2$, so it does not improve the
accuracy of the PT series compared to the three-loop one.}
 \label{fig:s_PT}
\end{figure}

To test that,  we present in Fig.~\ref{fig:s_PT} the relative
contributions of separate terms in the four-loop expansion
(\ref{Delta-ex})
\begin{equation}
{\rm N}_i(Q^2)={\delta_i (Q^2)}/{\Delta_{\rm Bj}(Q^2)}.
\end{equation}
\begin{figure}[h!]
 \centerline{\epsfig{file=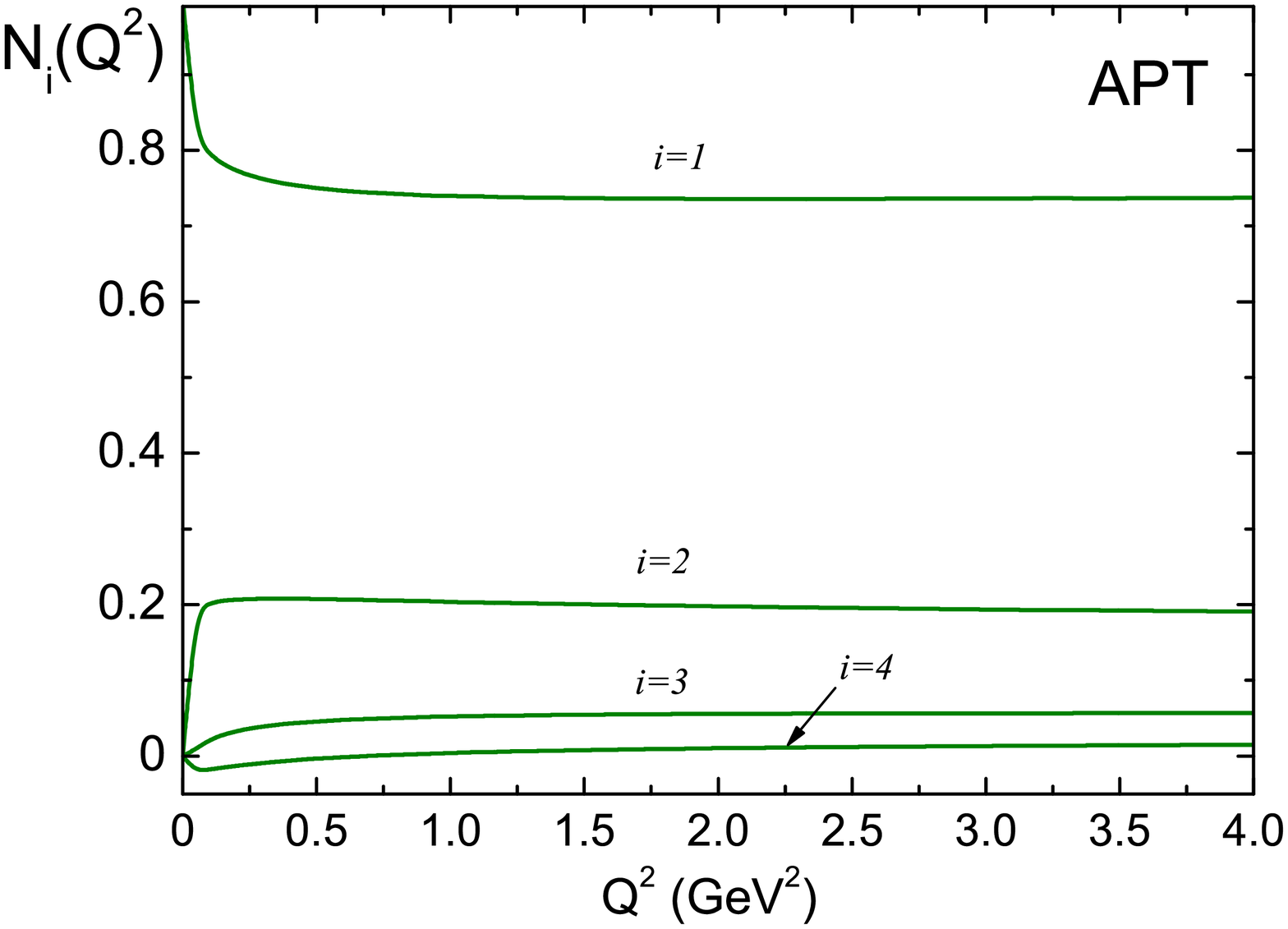,width=10cm}}
 \caption{The $Q^2$-dependence of the relative contributions of the
perturbative expansion terms in Eq.~(\ref{Delta_APT-Bj}) in the APT
approach. Third and fourth order contributions amount to less then 5
\% total, so the NLO APT approximation is sufficient for description
of the low energy JLab data at the current level of experimental
accuracy.}
 \label{fig:s_APT}
\end{figure}

As it is seen from Fig.~\ref{fig:s_PT}, in the region $Q^2<1$ GeV$^2$
the dominant contribution to the pQCD correction $\Delta_{\rm
Bj}(Q^2)$ comes from the four-loop term $\sim\as^4$. Moreover, its
relative contribution increases with decreasing $Q^2$. In the region
$Q^2>2$ GeV$^2$ the situation changes -- the major contribution comes
from one- and two-loop orders there. Analogous curves for the APT
series given by Eq.~(\ref{Delta_APT-Bj}) are presented in
Fig.~\ref{fig:s_APT}.

Figures~\ref{fig:s_PT} and \ref{fig:s_APT} demonstrate the essential
difference between the PT and APT cases, namely, the APT expansion
obeys much better convergence than the PT one. In the APT case, the
higher order contributions are stable at all $Q^2$ values, and the
one-loop contribution gives about 70 \%, two-loop -- 20 \%,
three-loop -- not exceeds 5\%, and four-loop -- up to 1 \%.

One can see that the four-loop PT correction becomes equal to the
three-loop one at $Q^2=2$ GeV$^2$ and noticeably overshooting it
(note that the slopes of these contributions are quite close in the
relatively wide $Q^2$ region) for $Q^2 \sim 1$ GeV$^2$ which may be
considered as an extra argument supporting an asymptotic character
of the PT series in this region.

In the APT case, the contribution of the higher loop corrections is
not so large as in the PT one. The four-loop order in APT can be
important, in principle, if the theoretical accuracy to better than
1 \% will be required.

\subsection{The $\mu$-scale dependence}

As it is known, any observable obtained to all orders in pQCD
expansion should be independent of the renormalisation scale $\mu$,
but in any truncated-order perturbative series the cancelation is
not perfect, such that the pQCD predictions depend on the choice of
the $\mu$-scale (for a review see, e.g., Ref.~\cite{Bethke:2009}).

In order to estimate this dependence of $\Gamma^{p-n}_1$ on the
unphysical re\-normalization-scale parameter $\mu$, we use the
four-loop expression for the coefficient function $C_{\rm
Bj}(\mu^2/Q^2)$ recently published in Ref.~\cite{Baikov:2010}.
One commonly introduce the dimensionless parameter $x_{\mu}$
($\mu^2=x_\mu Q^2$), which we have chosen to change within the
interval $x_{\mu} =  0.5\div2$ (see, for example, the analysis in
Ref.~\cite{Ch-tau}), and compare the $\mu$-scale ambiguities between
the three- and four-loop PT  series.

\begin{figure}[h!]
\centerline{\epsfig{file=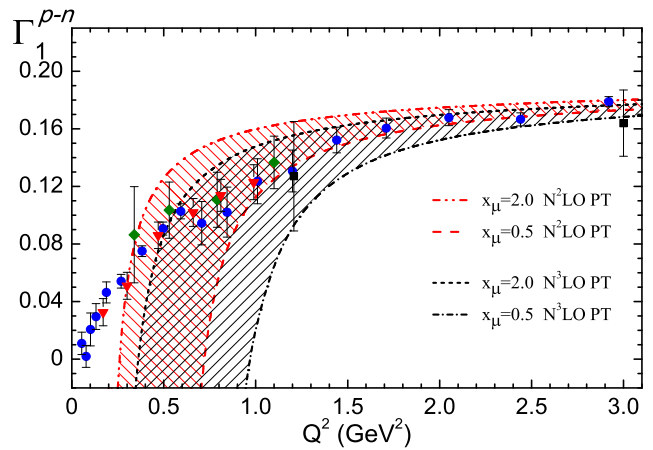,width=10.0cm}} \caption{The
$\mu$-scale ambiguities for the perturbative part of the BSR versus
$Q^2$ for three- (shaded region between dash-dot-dotted and dashed
lines) and four-loop (shaded region between short-dashed and
dash-dotted lines) orders of PT series corresponding to changing of
$x_{\mu}$ in the interval $0.5\div2$. These two regions have similar
widths and only slightly shifted w.r.t. each other, so the
differences between three- and four-loop results are within the
experimental error bars. Hence, in the common PT case, the N$^3$LO
approximation does not improve the data description compared to the
N$^2$LO one (see also Fig.~\ref{fig:s_PT}).}
\label{fig:Fig-mu_Q2}
\end{figure}

In Fig.~\ref{fig:Fig-mu_Q2}, the perturbative part of the BSR is
plotted as a function of $Q^2$ in three- and four-loop orders of PT
series corresponding to $x_{\mu}$ in the interval $0.5\div2$. The
width of the arising strip for the four-loop approximation is
similar to the one for the three-loop approximation in the highest
JLab region $Q^2 \sim 3 $ GeV$^2$ \footnote{One can find that an
account for four-loop contribution leads to a decrease of the
$\mu$-dependence if $Q^2 \geq 5 $ GeV$^2$ which is currently outside
the JLab kinematical range, but will be accessible by JLab after the
scheduled upgrade.}, so these approximations provide the description
of the data with comparable accuracy, as discussed above. Thus, the
four-loop result does not improve the data description noticeably in
the low-energy domain.

At the same time, for $Q^2 \leq 1 $ GeV$^2$, where PT does not allow
the description of the data, the inclusion of the four-loop
contribution leads to a stronger $\mu$-dependence.

These observations provide yet other arguments supporting the
mentioned transition to asymptotic PT series at $Q^2 \sim 1 $
GeV$^2$.

\section{Higher twists contribution}

\subsection{The results of the fit}

Now, using  expression~(\ref{PT-Bj-HT}) fitted to the above
mentioned experimental data~\cite{JLab08data,JLab-old-data} we
extract the coefficients $\mu_{2i}$ of the higher twist OPE
corrections. The minimal borders of fitting domains in $Q^2$ are
settled from the {\it ad hoc} restriction $\chi^2<1$ and monotonous
behavior of the resulting fitted curves.

Previously, a detailed higher-twist analysis of the two- and
three-loop expansions in powers of $\as$ was performed in
Refs.~\cite{Bjour}. Now, we extend the analysis up to an order
$\sim\alpha_s^4$.
\begin{figure}[h!]
 \centerline{\epsfig{file=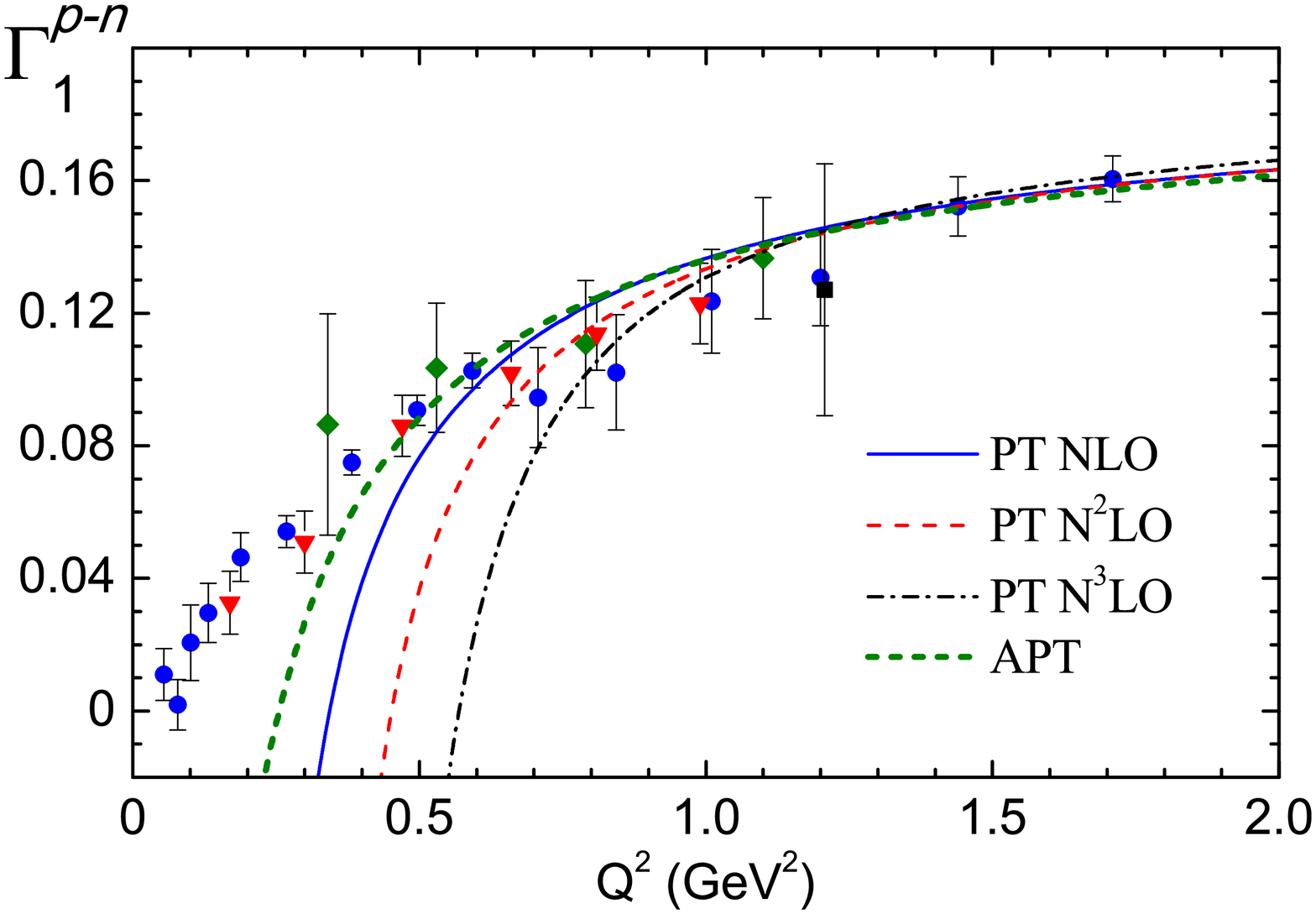,width=10cm}}
 \caption{The one-parametric $\mu_4$-fits of the BSR JLab data in
various (NLO, N$^2$LO, N$^3$LO) orders of the PT and the all-order
APT expansions. In the PT case, the four-loop result does not
improve the data description compared to the three-loop one. In the
APT case, the NLO approximation is sufficient due to higher-loop
stability of the APT expansion (see also Fig.~\ref{fig:s_APT}).}
 \label{fig:fit1}
\end{figure}

\begin{figure}[h!]
 \centerline{\epsfig{file=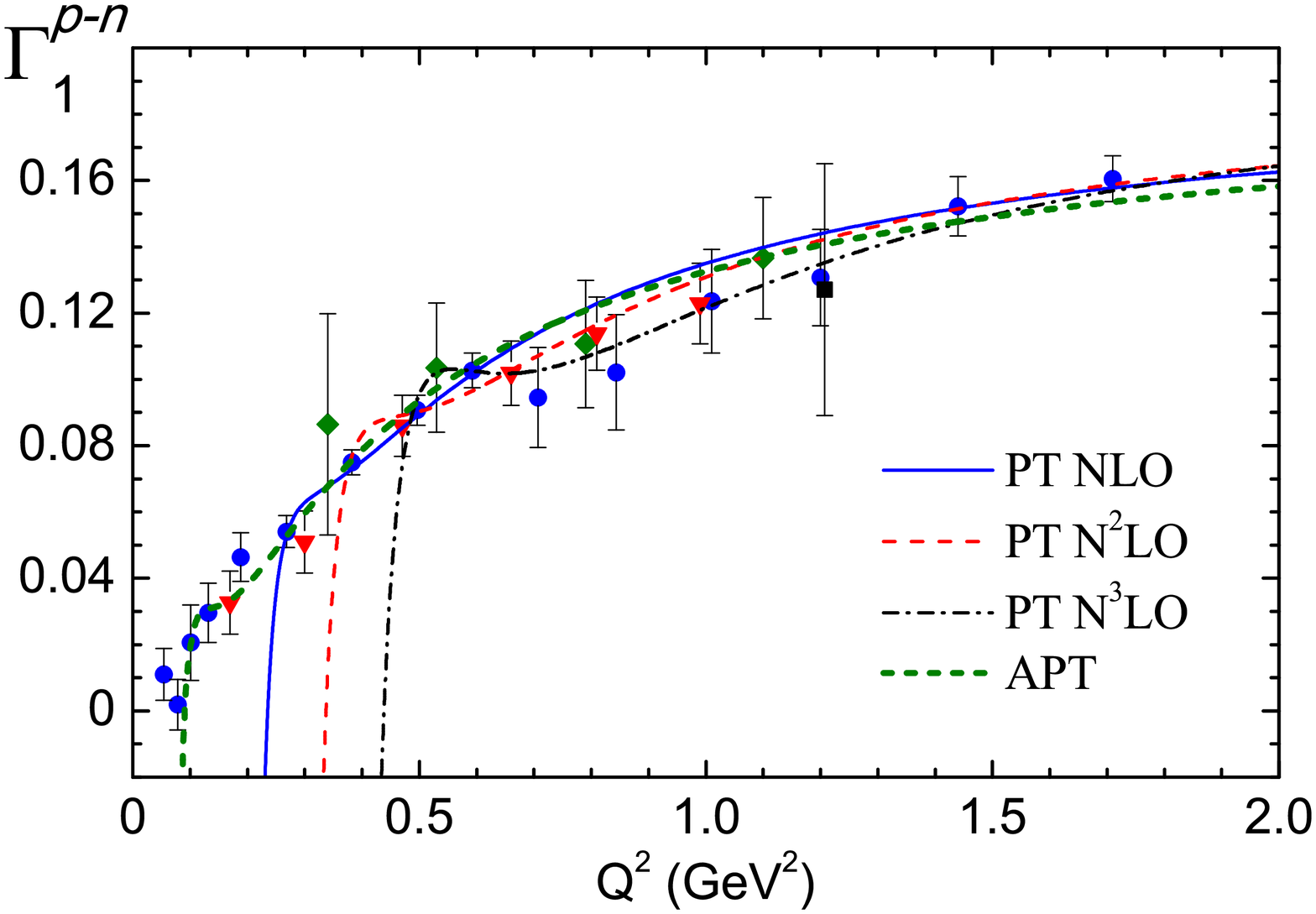,width=10cm}}
 \caption{The three-parametric $\mu_{4,6,8}$-fits of the BSR JLab data in
various (NLO, N$^2$LO, N$^3$LO) orders of the PT and the all-order
APT expansions.}
 \label{fig:fit2}
\end{figure}

In Figs.~\ref{fig:fit1} and~\ref{fig:fit2} we present the results of
1- and 3-parametric fits in various orders of PT and APT. The
corresponding fit results for  higher twist terms, extracted in
different orders of PT and APT, are given in
Table~\ref{tab:Bjtot_PT} (all numerical results are normalized to
the corresponding powers of the nucleon mass $M$). From these
figures and Table~\ref{tab:Bjtot_PT} one can see that APT allows one
to move down to $Q^2 \sim 0.1$  GeV$^2$ in description of the
experimental data \cite{Bjour}. At the same time, in the framework
of the standard PT the lower border shifts up to higher $Q^2$ scales
when increasing the order of PT expansion. This is caused by extra
unphysical singularities in the higher-loop strong coupling.
\begin{table}[!h] \small
\caption{Results of higher twist extraction from the JLab data on
BSR in various (NLO, N$^2$LO, N$^3$LO) orders of PT and all orders
of APT.}
\begin{center}\label{tab:Bjtot_PT}
\begin{tabular}{|l|l|c|c|c|} \hline
 Method& $Q_{min}^2,\,$ & $\quad\mu_4/M^2$ &
 $\quad\mu_6/M^4$ & $\quad\mu_8/M^6$  \\  \hline
 \multicolumn{5}{|c|}{The best $\mu_4$-fit results}\\ \hline
 PT NLO          &  ~$0.5$   & $-0.028(5)$  & $ - $        &  $ - $       \\
 PT N$^2$LO      &  ~$0.66$  & $-0.014(7)$  &   $ -  $     &   $ - $      \\
 PT  N$^3$LO     &  ~$0.71$  & $~~0.006(9)$ &    $ -$      & $-$  \\
  APT           & $~0.47$ & $-0.050(4)$ & $-$ & $-$  \\  \hline
 \multicolumn{5}{|c|}{The best $\mu_{4,6,8}$-fit results}  \\  \hline
PT  NLO              & $~0.27$ & $-0.03(1)$ & $-0.01(1)$   & $0.008(4)$   \\
PT N$^2$LO          &  ~$0.34$  & $~~0.01(2)$  & $-0.06(4) $  & $0.04(2)~$   \\
PT N$^3$LO          &  ~$0.47$ & $~~0.05(4)$  &  $-0.2(1)~$  & $0.12(6)~$    \\
APT          & $~0.08$ & $-0.061(4)$ & $0.009(1)$  & $-0.0004(1)$  \\ \hline
\end{tabular} \end{center}
\end{table}

\subsection{Sensitivity of the higher twists to $\Lambda_{\rm {QCD}}$ variations}

In the above analysis, we normalized $\as$ at the $Z$-boson mass
scale and then fixed the value of the $\Lambda$ parameter separately
in each order in $\as$ approximation (it was sufficient for
understanding the role of the fourth order in the PT/APT
perturbative series). However, the corresponding values of the
$\Lambda$ parameter extracted in this way may be different from ones
obtained in the direct QCD analysis of the experimental data on the
moments of the structure functions (see, e.g.,
Ref.~\cite{Leader:2010rb}).
Having this in mind, we investigate additionally the sensitivity of
the extracted values of the higher twist term $\mu_4$ to the QCD
scale parameter $\Lambda$ in various orders of PT. In the framework
of APT, the sensitivity of $\mu_4$ to the $\Lambda$ parameter is
weak, and it does not depend on the order of the loop expansion.
Correspondingly, the values of the higher twist coefficients turn
out to be considerably more precise than those extracted in the PT
approach (see also Table~\ref{tab:Bjtot_PT}).

\begin{figure}[h!]
  \centerline{\epsfig{file=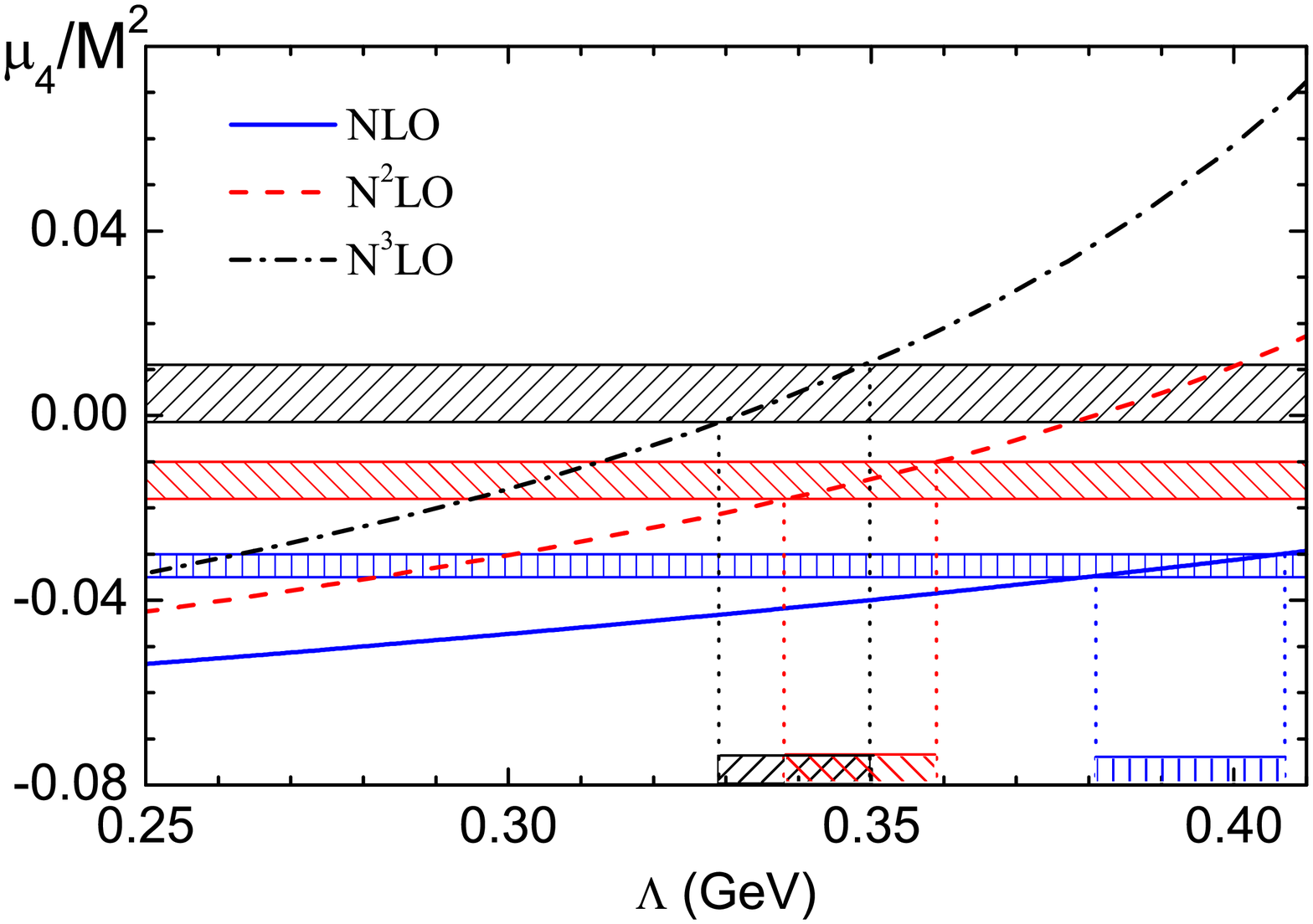,width=10.0cm}}
\caption{Value of the higher twist coefficient $\mu_4$ extracted
from the JLab data using the PT at different orders at
$Q_{min}^2=0.66$~GeV$^2$ with error bands. Vertical lines denote the
corresponding uncertainty ranges in $\Lambda$-parameter. The ranges
corresponding to N$^2$LO and N$^3$LO approximations have similar
sizes and overlap with each other, so the four-loop result does not
improve the stability w.r.t. $\Lambda$ variations compared to the
three-loop one.}
 \label{fig:mu4-chi}
\end{figure}

In Fig.~\ref{fig:mu4-chi} we show values of the coefficient $\mu_4$
extracted from the JLab data using two-, three- and four-loop PT at
$Q_{min}^2=0.66$~GeV$^2$ vs the parameter $\Lambda$. One can see
that the PT  does not lead to a stable result for extracted $\mu_4$
value with respect to $\Lambda$ variations. The extracted higher
twist coefficient $\mu_4$ changes quite strongly between different
orders of the PT expansion. And it happens in both in absolute value
and sign, namely, at $\Lambda>320$~MeV the higher twist coefficient
becomes positive in the four-loop PT order. This  sensitivity of the
higher twist term $\mu_4$ to variations of the $\Lambda$  becomes
stronger at higher PT orders.

On the other hand, these data tell us that the absolute value of
$\mu_4$  decreases with the order of PT and just at four-loop order
becomes compatible to zero. This may be considered as a
manifestation of duality between higher orders of PT and HT (see
Ref.~\cite{Bjour} and references therein). Moreover, when PT series
manifests the asymptotic behavior (i.e. becomes most close to exact
result), the HT (which may be considered as a contribution
completing the PT series)  can be reduced to zero.

\section{Summary and Conclusion}

In this work, we performed the QCD analysis of the precise low
energy JLab data on the BSR in the N$^3$LO PT order and extracted
the OPE higher twist terms using the four-loop expression for the
QCD correction to the Bjorken integral $\Delta_{\rm Bj}$ published
recently in Ref.~\cite{Baikov:2010}.

Our main observations are:

i) The four-loop approximation provides good description of the data
for the highest JLab $Q^2 \sim 3$ GeV$^2$. For several data points
there is an impression that the four-loop approximation is better
than the three-loop one. At the same time, the order of magnitude of
both these contributions is the same as an experimental error, so a
more precise statement can hardly be made.

ii) For lower $Q^2 \leq 0.7 $ GeV$^2$ the four-loop PT contribution
 does not help to describe the data. Meanwhile, as it was shown
 earlier \cite{Bjour}, the APT application leads to higher loops
 stability of the HT extraction. In turn, this results in accurate
 data description down to $Q^2\sim 0.1$ GeV$^2\,$ always at the
 two-loop APT level (see Fig.~\ref{fig:fit2}).

iii) The magnitude of HT decreases with an order of PT and becomes
compatible to zero at the four-loop level.
\bigskip

Our concluding impression is that all these features may indicate
that the asymptotic nature of the QCD PT series is revealed at the
four-loop level at $Q^2 \sim 1 $GeV$^2$. This conjecture is
confirmed by the analysis of relative contributions of various PT
terms, as well as by that of unphysical $\mu$-dependence.



\section*{Acknowledgments}

We are thankful to S.V.~Mikhailov and K.G.~Chetyrkin for valuable
discussions as well as to A.V.~Sidorov and D.B.~Stamenov for useful
comments and to V.V.~Skalazub for interest in the work and
stimulating discussions.

 This work was partly supported by the Russian presidential grant
 Scient. School--3810.2010.2, the RFBR grants
 09-02-00732, 09-02-01149, 11-01-00182, the BelRFFR-–JINR grant F10D-001, and
 by the Carl Trygger Foundation.

\end{document}